\documentclass[reprint,amsmath,amssymb,aps]{revtex4-2}
\usepackage{graphicx}
\usepackage{dcolumn}
\usepackage{bm}
\usepackage{xcolor}
\usepackage{float}
\usepackage{mathrsfs}
\usepackage{lipsum}

\begin{document}

\preprint{APS/123-QED}

\title{Dynamical analysis of a parameter-aware reservoir computer}

\author{Dishant Sisodia}\email{dishantsisodia.physics@gmail.com}
\author{Sarika Jalan} \email{sarika@iiti.ac.in}
\affiliation{Complex Systems Lab, Department of Physics, Indian Institute of
Technology Indore, Khandwa Road, Simrol, Indore-453552, India}

\date{\today}

\begin{abstract}
Reservoir computing has been shown to be a useful framework for predicting critical transitions of a dynamical system if the bifurcation parameter is also provided as an input. Its utility is significant because in real-world scenarios, the exact model equations are unknown. This Letter shows how the theory of dynamical system provides the underlying mechanism behind the prediction.
Using numerical methods, by considering dynamical systems which show Hopf bifurcation, we demonstrate that the map produced by the reservoir after a successful training undergoes a Neimark-Sacker bifurcation such that the critical point of the map is in immediate proximity to that of the original dynamical system. In addition, we have compared and analyzed different structures in the phase space. Our findings provide insight into the functioning of machine learning algorithms for predicting critical transitions.
\end{abstract}

\maketitle

\paragraph{Introduction:} 

Machine learning is increasingly being utilized in various diverse areas of complex system research, such as the analysis of dynamical systems \cite{10.1063/5.0016505}, model free prediction of chaotic systems \cite{PhysRevLett.120.024102}, solving the Fokker-Planck equation \cite{10.1063/1.5132840}, fatigue detection using EEG datasets \cite{10.1063/1.5120538}, finding an optimal set of nodes in complex networks \cite{Fan2020}, etc. Understanding the mechanisms by which machines make accurate predictions from a dynamical perspective is an emerging research area. Using feed forward neural networks one recent study \cite{PhysRevLett.132.057301} has shown that the finite-time Lyapunov exponents form geometrical structures in the input space by segmenting it into distinct regions which the network then correlates with different classes. Thus, from a dynamical perspective, the exponential sensitivity of the output near the decision boundary was explained.

Reservoir computing (RC) \cite{LUKOSEVICIUS2009127,doi:10.1126/science.1091277} is a machine learning paradigm based on the recurrent neural network (RNN) framework, which has recently gained significant attention. In contrast to other machine learning methods, in RC only the final output weights are trained via simple linear regression keeping the hidden layer unchanged, thus making the training part very efficient. Therefore, RC becomes more favorable for sequential tasks such as chaotic time series forecasting \cite{10.1063/1.5039508,10.1063/1.4979665,PhysRevE.98.052209}, speech recognition \cite{Torrejon2017}. Another advantage of RC is its hardware implementation in a variety of physical systems \cite{TANAKA2019100}, which motivates the development of fast information processing devices. A recent study to examine the mechanism of RC \cite{PhysRevE.105.L052201} indicates that successful training leads to synchronization of reservoir oscillators into clusters which obey a power law. 
\begin{figure}[t!]
\includegraphics[scale=0.4]{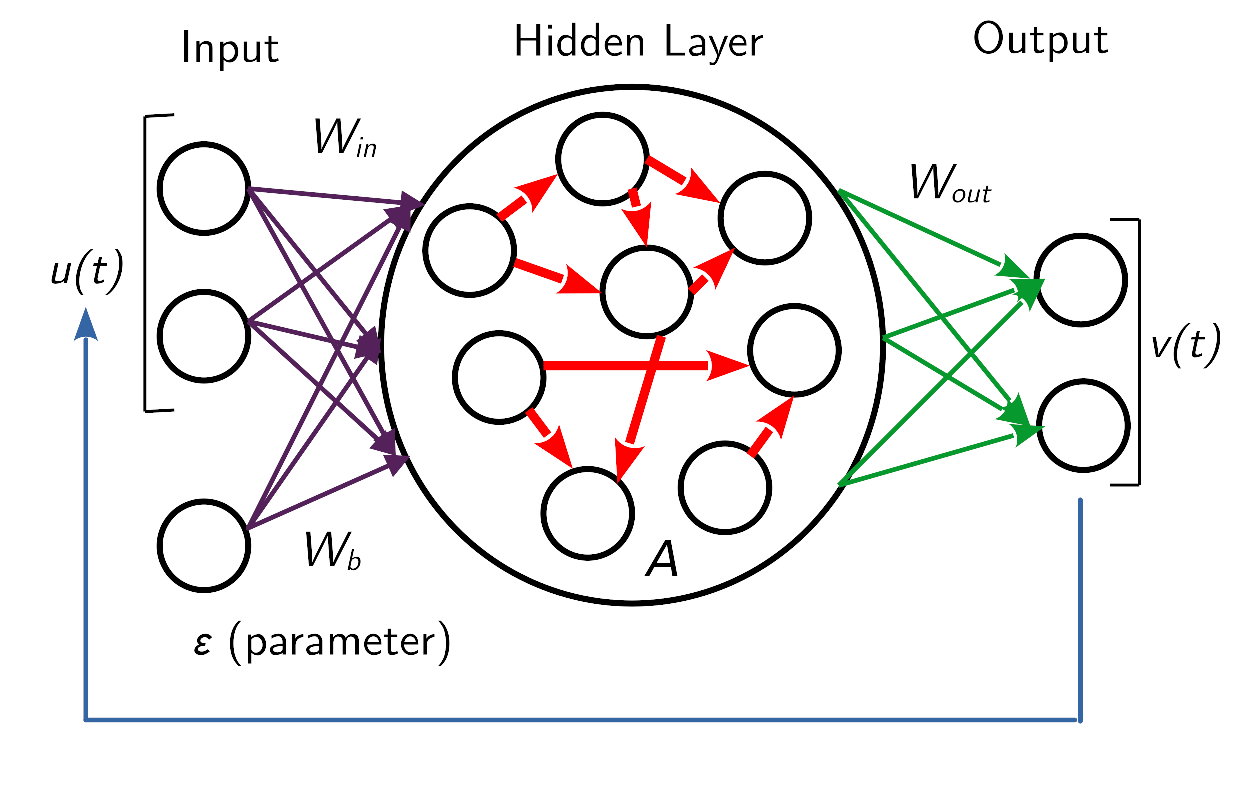}
\caption{Schematic diagram of parameter-aware architecture having an additional input parameter channel $\varepsilon$. $W_{in}, W_b$ are the weight matrices for input $u$ and $\varepsilon$, respectively. $\mathcal{A}$ is the adjacency matrix of the reservoir. $W_{out}$ is determined after training and during the testing phase, current output serves as the input for the next time step. }
\label{fig1}
\end{figure}

Furthermore, by incorporating the system parameter as an additional input to RC, known as the ``Parameter-aware architecture", it has been demonstrated that it is possible to train a system for a set of different values of the bifurcation parameter before the critical transition and correctly predict the value of the parameter at which the transition occurs \cite{PhysRevE.104.014205,PhysRevResearch.3.013090}. Moreover, the framework replicates the dynamics of the original system in the vicinity of the transition point. It becomes immensely useful because, in the real world, the exact system equations are often unknown. Parameter(s) of a system might be slowly drifting due to external factors inducing a bifurcation which may lead to undesirable phenomenon such as amplitude death \cite{SAXENA2012205, KOSESKA2013173} or a crisis \cite{GREBOGI1983181}. Therefore, understanding the underlying mechanism is crucial for advancing this domain. To our knowledge, no studies have so far discussed the operational mechanism of a parameter-aware RC model. This Letter uses bifurcation analysis to investigate this mechanism.

\paragraph{Model:}
A reservoir computing machine projects an $n$ dimensional input channel $u(t)$ into a higher $m$ dimensional space through a weight matrix $W_{in}$. The adjacency matrix $(m \times m)$ of the reservoir network $\mathcal{A}$ is responsible for including past memory of the reservoir state. The matrices $W_{in}$ and $\mathcal{A}$ are initially chosen and are fixed throughout the training process. $W_{in}$ is chosen from a uniform random distribution ranging from $[-b,b]$. Often, $\mathcal{A}$ is chosen to be the adjacency matrix of an Erd\"os–R\'enyi model network \cite{doi:10.1126/science.286.5439.509}, having a connection probability $\sigma$ and scaling $\rho$. For a given value of spectral radius, $\mathcal{A}$ is scaled such that its largest eigenvalue is $\rho$. The reservoir state $r(t)$ is updated by the following equation for $N_t$ time steps:
\begin{equation}
    r[i+1] = (1-\alpha)r[i] + \alpha \hspace{0.1cm} tanh(A r[i] + W_{in} u[i+1]),
\end{equation}
where $\alpha$ is known as the leakage rate. The updated reservoir state is influenced both by the previous reservoir state and the current input state with $\alpha$ dictating the relative dominance of these two factors in determining the new reservoir state.
During the training phase, the aim is to minimize the difference between the actual and predicted output. To achieve this, the updated reservoir states are stored and stacked to form a matrix $\mathcal{R}$ of dimensions $(m \times N_t)$. The actual output which is the target variable is also stacked to form a matrix $\mathcal{U}$ of dimensions $(n \times N_t)$. The output weight matrix is then calculated using Tikhonov regularization:
\begin{equation}
    W_{out} = \mathcal{U}.\mathcal{R}^{T}.(\mathcal{R}\mathcal{R}^{T} + \beta \mathcal{I})^{-1}.
\end{equation}
\begin{figure}[t!]
    \includegraphics[width = 0.43\textwidth]{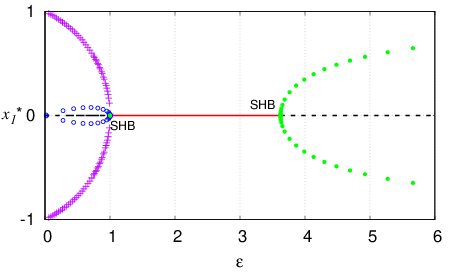}
    \caption{Bifurcation plot of coupled Stuart-Landau system. Solid green, hollow blue dots represent stable, unstable limit cycle, respectively. Solid red, dashed black line represent stable and unstable fixed point, respectively. Purple crosses indicate the torus part. SHB represents supercritical Hopf bifurcation. }
    \label{fig:2}
\end{figure}
In the testing phase,  the output of the current state acts as the input for the next state and the output $v(t)$ is given by $W_{out} r(t)$.

As illustrated in Fig.~\ref{fig1}, the  ``Parameter-aware architecture'' which has been used in previous studies has a modification. There is an additional input channel connected to the reservoir which has the information of the parameter. Therefore, the dynamical evolution of the reservoir states \cite{PhysRevResearch.3.013090} is described as follows:
\begin{equation}\label{eq1}
\begin{aligned}
    r[i+1] =& (1-\alpha)r[i]  +  \alpha \hspace{0.1cm} tanh(A r[i] + \\&W_{in} u[i+1] + k_{b} W_b (\varepsilon - \varepsilon_b)).
\end{aligned}
\end{equation}
Here, $\varepsilon$ is the bifurcation parameter, $k_b$ and $\varepsilon_b$ are the hyperparameters. $W_b$, initially chosen from a uniform random distribution ranging from $[-b,b]$, remains fixed throughout the training. This gives us the following set of hyperparameters which can be tuned and optimized, $(m, b,\alpha, \rho, \sigma, k_b, \varepsilon_b)$. RC is trained for multiple values of $\varepsilon$ before the transition point. After training, it can predict the dynamics correctly in the vicinity of the transition point (both before and after the transition point).
\begin{figure}[t!]
\begingroup
\begin{tabular}{c}
\includegraphics[width = 0.23\textwidth,height=4cm]{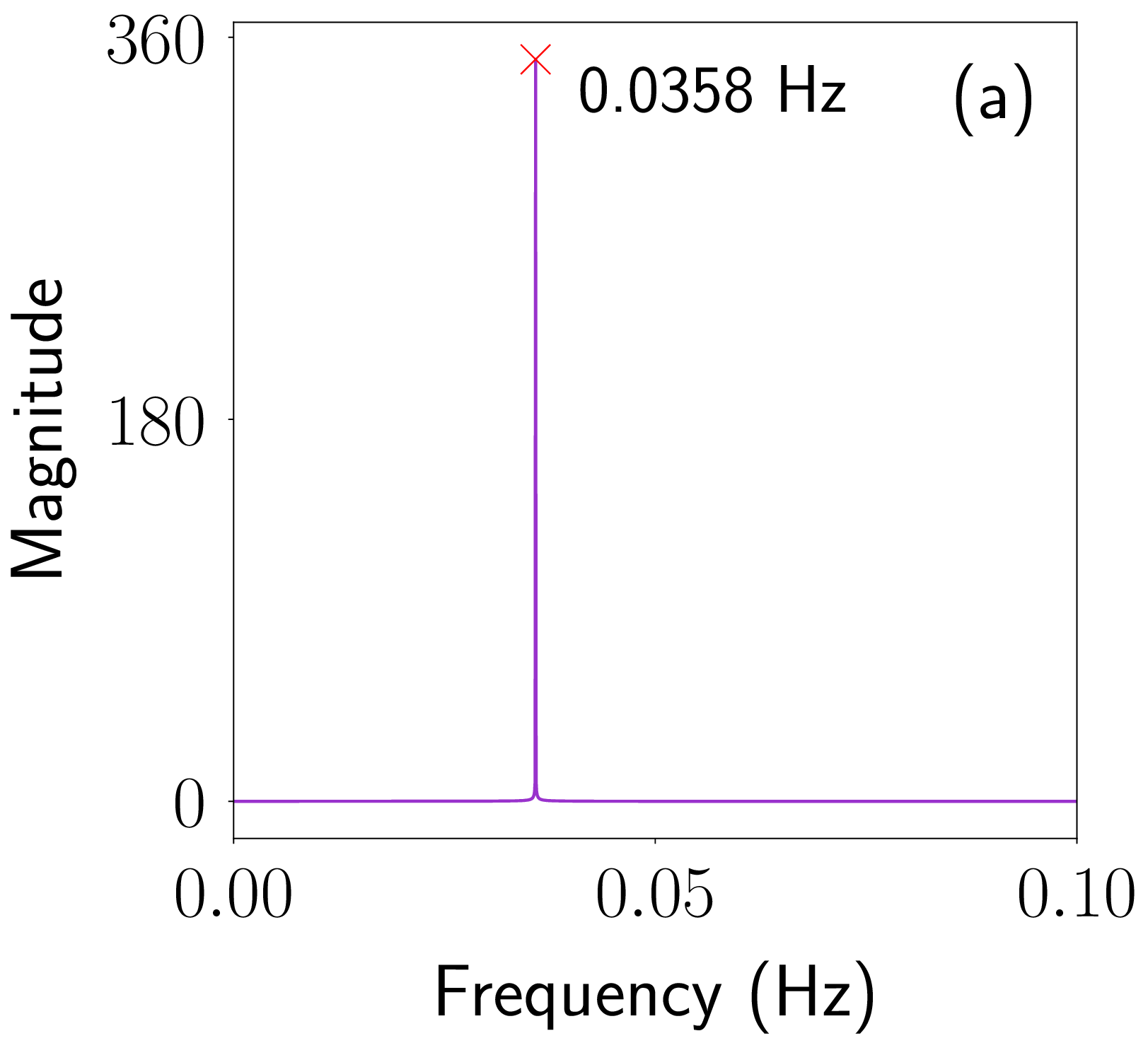}
\includegraphics[width = 0.23\textwidth,height=4cm]{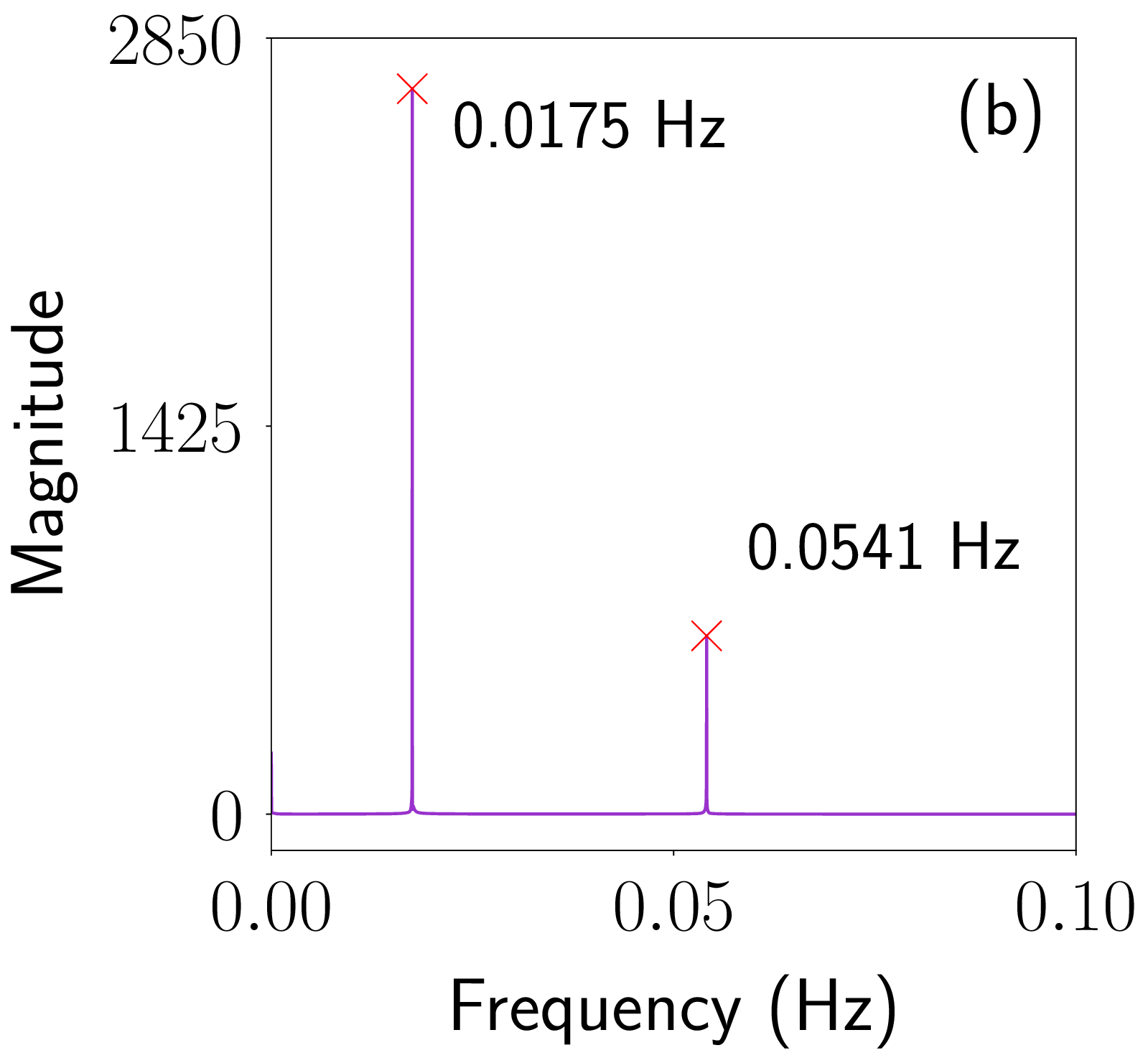}

\end{tabular}
\endgroup
\caption{The dominant frequencies of the time series are isolated along with their magnitudes by performing the Fourier transform of $x_1(t)$ for coupled Stuart-Landau oscillator at (a) $\varepsilon = 3.63$, (b) $\varepsilon = 0.99$. }
\label{fig:3}
\end{figure}

\paragraph{Results:}
During the testing phase, the output of the current state acts as the input of the next state. Using this information, we can modify Eq.~\ref{eq1} to obtain:
\begin{equation} \label{eq2}
\begin{aligned}
    r[i+1] =& (1-\alpha)r[i]  +  \alpha \hspace{0.1cm} tanh(A r[i] + \\&W_{in}W_{out} r[i] + k_{b} W_b (\varepsilon - \varepsilon_b)),
\end{aligned}
\end{equation}
which can be further simplified to a simple map equation;
\begin{equation} \label{eq3}
    r[i+1] = (1-\alpha)r[i]  +  \alpha \hspace{0.1cm} tanh(\Lambda r[i] + \Omega).
\end{equation}
This forms a system of $m$ autonomous map equations, where $m$ is the dimension of the reservoir and $r[i]$ and $\Omega$ both are $m$ dimensional column vectors, while $\Lambda$ is a $m \times m$ matrix. The matrix $\Lambda$ is ($\mathcal{A} + W_{in}W_{out}$) and $\Omega$ equals ($k_{\varepsilon} W_b (\varepsilon - \varepsilon_b)$). A successful prediction implies that the learned weight matrix $W_{out}$ makes the map in Eq.~\ref{eq3} to mimic the bifurcation pattern of the original continuous dynamical system. The final output which generates a time series at the desired value of the bifurcation parameter is just a linear combination of reservoir states ($v[i] = W_{out} r[i]$). This implies that a thorough study of  Eq.~\ref{eq3} will explain the final output. Here, we first numerically compute the fixed point $r^{*}$ of Eq.~\ref{eq3} and then evaluate the Jacobian matrix $\mathcal{J}$ at $r^*$ as follows:
\begin{equation} \label{eq4}
\begin{aligned}
    \delta r[i+1] &= (1-\alpha)\delta r[i] + \{\mathbb{I} - tanh^{2}(\Lambda r^{*} + \Omega)\}\Lambda \delta r[i] \\&
    = \mathcal{J} \delta r[i].
\end{aligned}
\end{equation}
Behaviour of the eigenvalues of $\mathcal{J}$ with changing $\varepsilon$ is then analyzed. We demonstrate the results by considering two examples, coupled Stuart-Landau oscillators and Van der Pol oscillators.

\paragraph{Coupled Stuart-Landau oscillators:}
The four dimensional model of two coupled, non identical Stuart-Landau oscillators with diffusive coupling \cite{ARONSON1990403} can be described by the following set of equations:
\begin{eqnarray} \label{eq5}
    &&\dot x_{i} = (1 - x_{i}^{2} - y_{i}^{2}) x_{i} - \omega_{i} y_{i} + \varepsilon (x_j - x_i)\\
    &&\dot y_{i} = (1 - x_{i}^{2} - y_{i}^{2}) y_{i} + \omega_{i} x_{i} + \varepsilon (y_j - y_i).
\end{eqnarray}
Here, we fix $\omega_1 = 2$ and $\omega_2 = 7$. The detailed bifurcation structure of the system is analyzed using the XPPAUT package \cite{10.1115/1.1579454} as depicted in Fig.~\ref{fig:2}. 

There exists a stable fixed point for $1< \varepsilon < 3.6$ which loses its stability through a hopf bifurcation and a stable limit cycle appears for $\varepsilon > 3.6$. Since the initialization of matrices $W_{in}, W_{b}, \mathcal{A}$ is random, after training, a different map is generated each time. However, we find that upon a successful training (if the actual and the predicted outputs match), the corresponding bifurcation point of the produced map always comes out to lie in a close proximity to that of the original system. This analysis is performed for one such random initialization of $W_{in}, W_{b}, \mathcal{A}$. The dimension of reservoir is chosen to be 300. After training the reservoir, the bifurcation occurs around $\varepsilon = 3.62$, which is remarkably close to the original system (Fig.~\ref{fig:5}).
\begin{figure}[t!]
\includegraphics[width=0.46\textwidth]{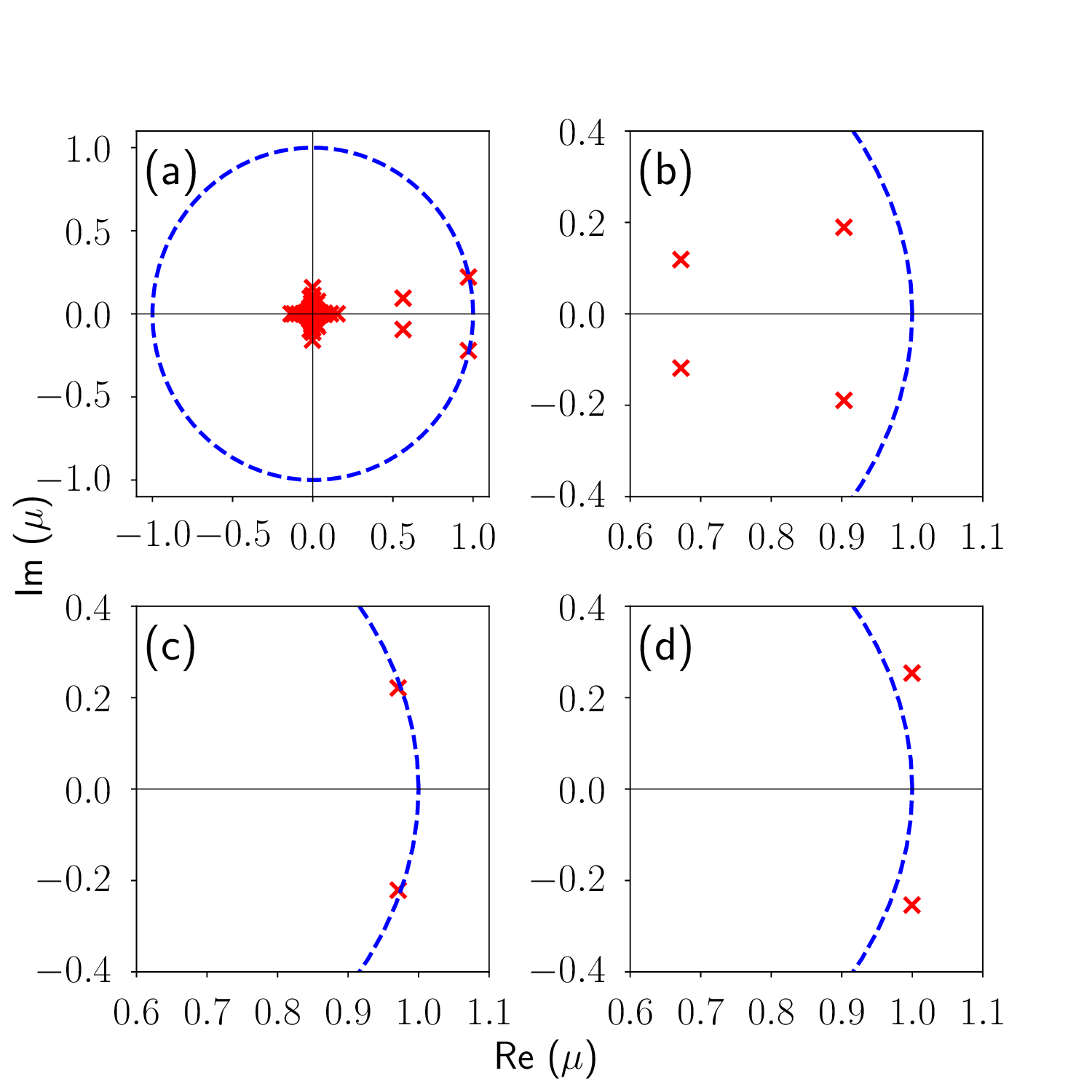}
    \caption{Behaviour of eigenvalues ($\mu$) of the map corresponding to the coupled Stuart-Landau oscillators with  change in $\varepsilon$ in the limit cycle regime. Unit circle in the complex plane is depicted using blue dashed lines. Red crosses show the position of eigenvalues in the complex plane. (a) A single conjugate pair crosses the unit circle at  $\varepsilon = 3.62$.  (b), (c) and (d) show the position of eigenvalues at $\varepsilon = 3.35$, $3.62$ and $3.9$ respectively.}
    \label{fig:4}
\end{figure}
We analyze the eigenvalues of $\mathcal{J}$ in the complex plane in the neighborhood of the bifurcation parameter. Some of the eigenvalues are real, rest exist in complex conjugate pairs. Before $\varepsilon = 3.62$, all the eigenvalues of $\mathcal{J}$ evaluated at the fixed point lie inside a unit circle. This implies that the learned map produces a stable spiral in the $\varepsilon < 3.62$ regime. Since the final outputs ($x_{i}, y_{i}$) are a linear combination of the reservoir states $v[i] = W_{out} r[i]$, the time series of $x_{i}, y_{i}$ will also be a stable spiral mimicking the original system. 

\begin{figure}[t!]
\begingroup
\begin{tabular}{c}
\includegraphics[scale=0.3]{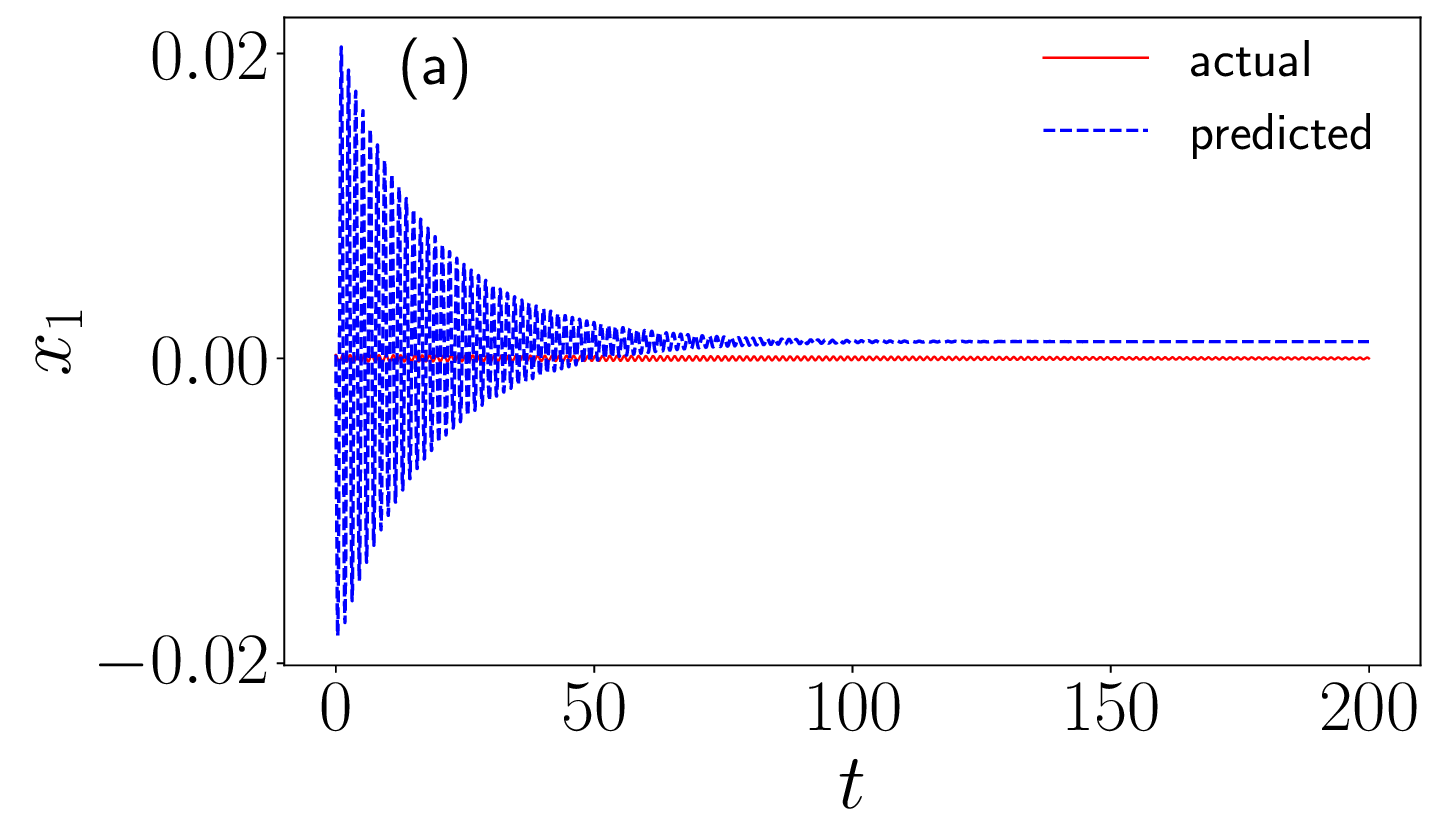}\\
\includegraphics[scale=0.3]{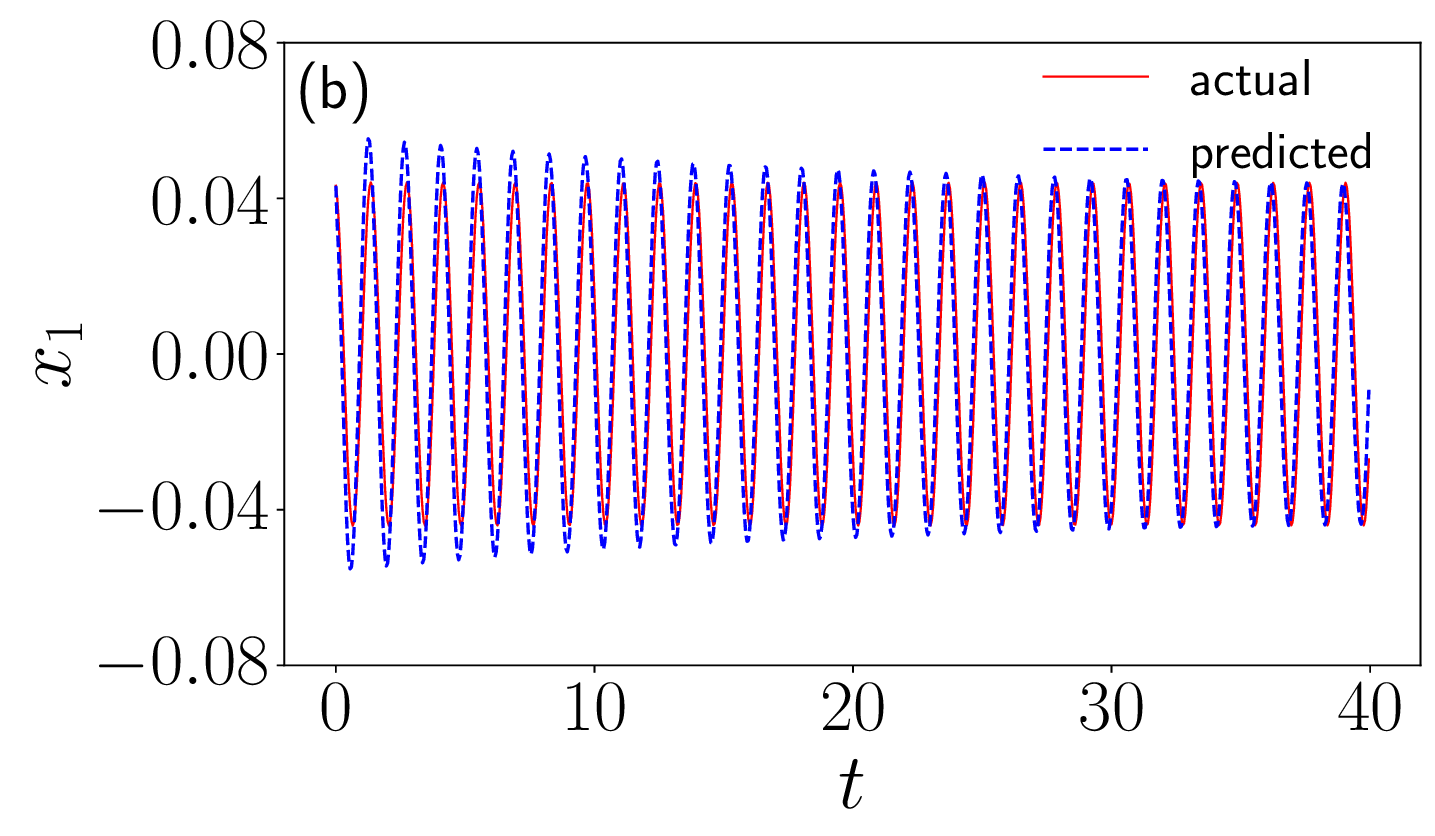}

\end{tabular}
\endgroup
\caption{Comparison of actual and predicted data of the coupled Stuart-Landau system for parameter value lying in the limit cycle regime, (a) $\varepsilon=3.61$, (b) $\varepsilon=3.63$. The chosen hyperparameters are ($m = 300, b = 2, \alpha = 1, \rho = 0.18, \sigma = 0.03, k_{\varepsilon} = 0.52, \varepsilon_b = 0$). The RC was trained at $\varepsilon = 3.75, 3.7, 3.65$.} 
\label{fig:5}
\end{figure}
In the $\varepsilon > 3.62$ regime, only one complex conjugate pair crosses the unit circle as the system transitions from the state of amplitude death to oscillations (Fig.~\ref{fig:4}). This is indicative of a Neimark-Sacker bifurcation. Further details on the Neimark-Sacker bifurcation can be found in the supplementary material \cite{supp} (see also references \cite{Kuznetsov:2008,Kuznetsov2023,Wiggins2003} therein). Here, all the $m = 300$ elements of the $r[i]$ column vector have a common period thus producing an invariant curve in the phase space formed by every $r_{j}, r_{k}$ pair. Therefore, all the output time series ($x_{i}, y_{i}$) which are again a linear combination of $r[i]$, also have the same common period of the oscillation. This explains how the map learns to correctly predict the dynamics in the oscillatory regime. It also elucidates the formation of an invariant ellipse in the $x_{i}, y_{i}$ phase plane, as illustrated in Fig.~\ref{fig:7}(b).
\begin{figure}[t!]
\begingroup
\begin{tabular}{c}
\includegraphics[scale=0.3]{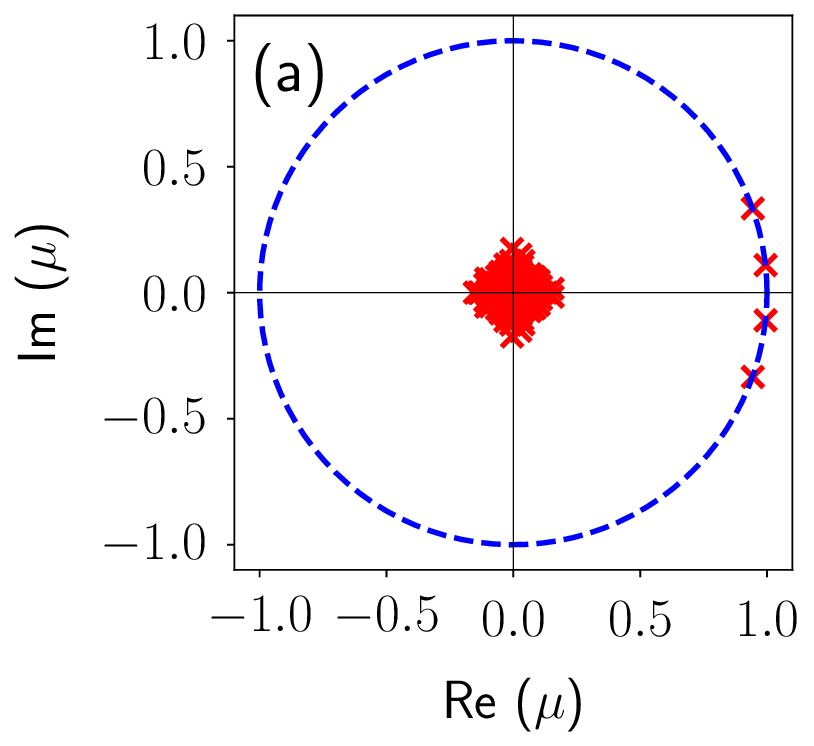}
\includegraphics[scale=0.3]{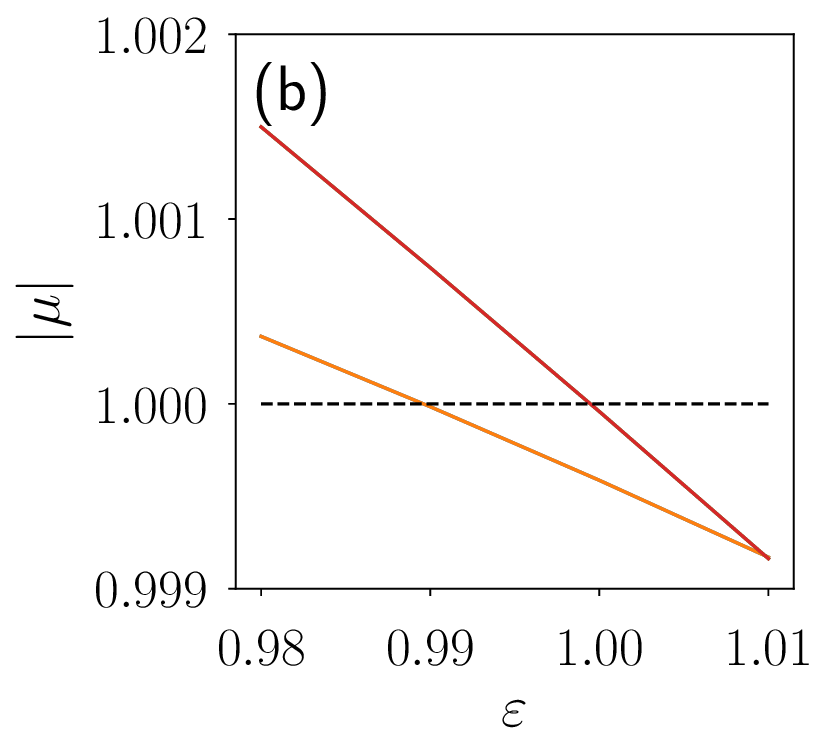}

\end{tabular}
\endgroup
\caption{Behaviour of eigenvalues ($\mu$) of the map corresponding to the coupled Stuart-Landau oscillators with  change in $\varepsilon$ in the torus regime. Unit circle is depicted using blue dashed line, and red crosses represent eigenvalues. Dashed black line corresponds to $|\mu| = 1$. (a) Two conjugate pairs crossing the unit circle at $\varepsilon = 1$. (b) Modulus of two conjugate pairs of eigenvalues $|\mu|$ plotted as a function of $\varepsilon$ depicting that the two pairs do not cross the unit circle at the same value of $\varepsilon$.}
\label{fig:6}
\end{figure}
\begin{figure}[t!]
\begingroup
\begin{tabular}{c c}
\includegraphics[scale=0.28]{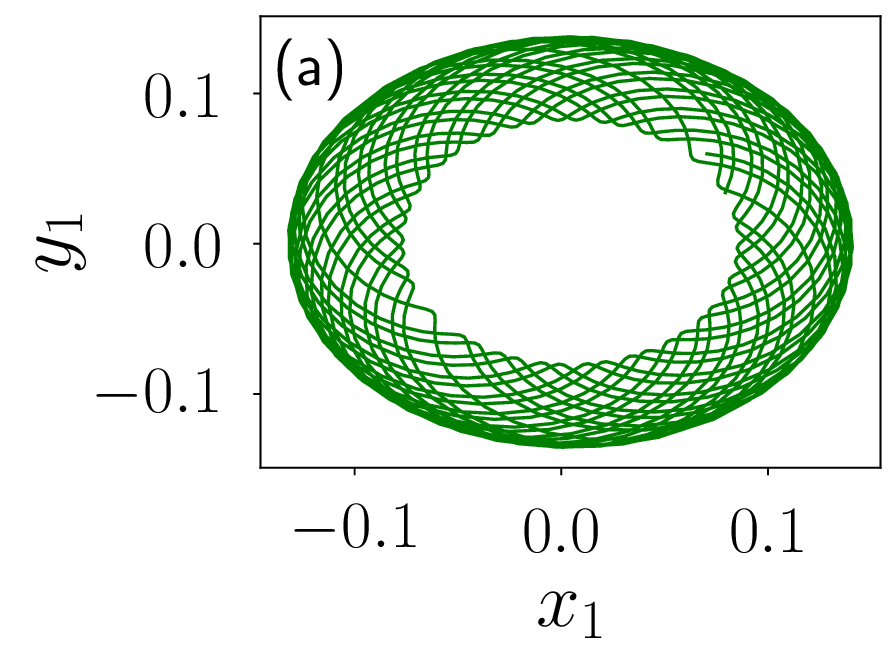}
\includegraphics[scale=0.28]{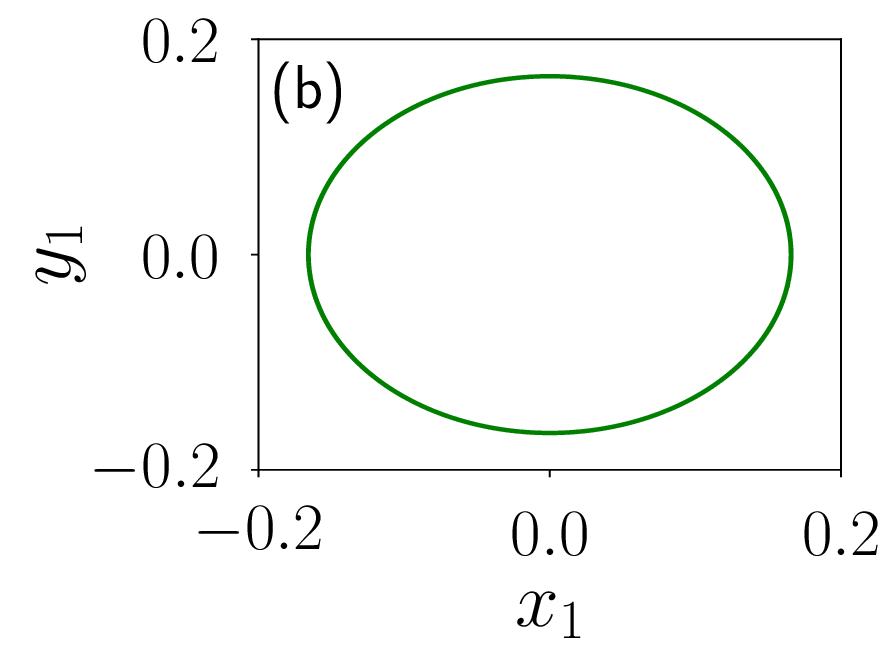}
\end{tabular}
\endgroup
\caption{Coupled Stuart-Landau oscillators system (a) Torus formation on the phase plane at $\varepsilon = 0.98$. (b) Invariant ellipse on the phase plane at $\varepsilon = 3.7$.}
\label{fig:7}
\end{figure}
The original system also has a bifurcation point around $\varepsilon = 1$. In this region, the phase space dynamics is very different from that in the $\varepsilon = 3.6$ case. As we reduce the value of $\varepsilon$, there exists a stable fixed point that loses its stability giving rise to a stable limit cycle at $\varepsilon = 1$. However, at the same point, a torus bifurcates from the stable limit cycle. Therefore, before $\varepsilon = 1$, a torus is produced in contrast to the closed invariant ellipse in the $x_1, y_1$ plane, which is obtained for the limit cycle after $\varepsilon = 3.6$ (Fig.~\ref{fig:7}).
After successfully training the system close to $\varepsilon = 1$ as depicted in Fig.~\ref{fig:8} we repeat the eigenvalue analysis. In this case, instead of a single pair crossing, around $\varepsilon = 1$, the corresponding generated map has two conjugate pairs crossings (Fig.~\ref{fig:6}). A single pair crossing generates a closed invariant curve, as witnessed for $\varepsilon = 3.6$. Thus, to correctly predict the torus, there arises a second pair crossing in the learned map model. Torus generation has been reported in the case of a double Neimark-Sacker bifurcation (a codimension-2 bifurcation) in which two pairs of conjugate eigenvalues cross the unit circle simultaneously \cite{LUO2006154,article}. However, since our system has a codimension-1 bifurcation (involving a single bifurcation parameter), at the bifurcation point, at most a single pair can cross the unit circle \cite{Kuznetsov2023}. Therefore, first a single pair crossing occurs via the usual Neimark-Sacker bifurcation followed by another conjugate pair crossing as depicted in Fig.~\ref{fig:6}.
\begin{figure}[t!]
\begingroup
\begin{tabular}{c}
\includegraphics[scale=0.3]{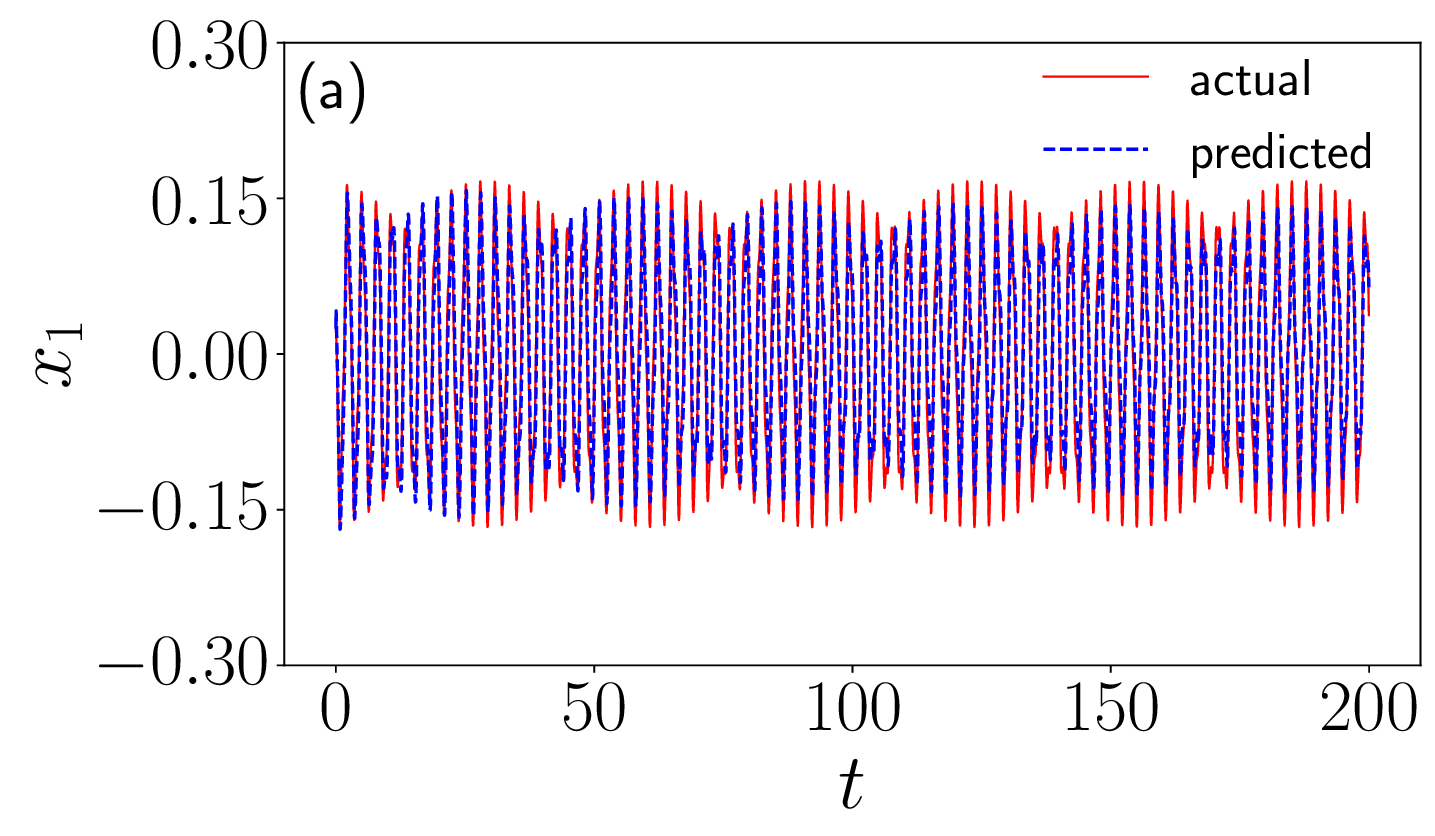}\\
\includegraphics[scale= 0.3]{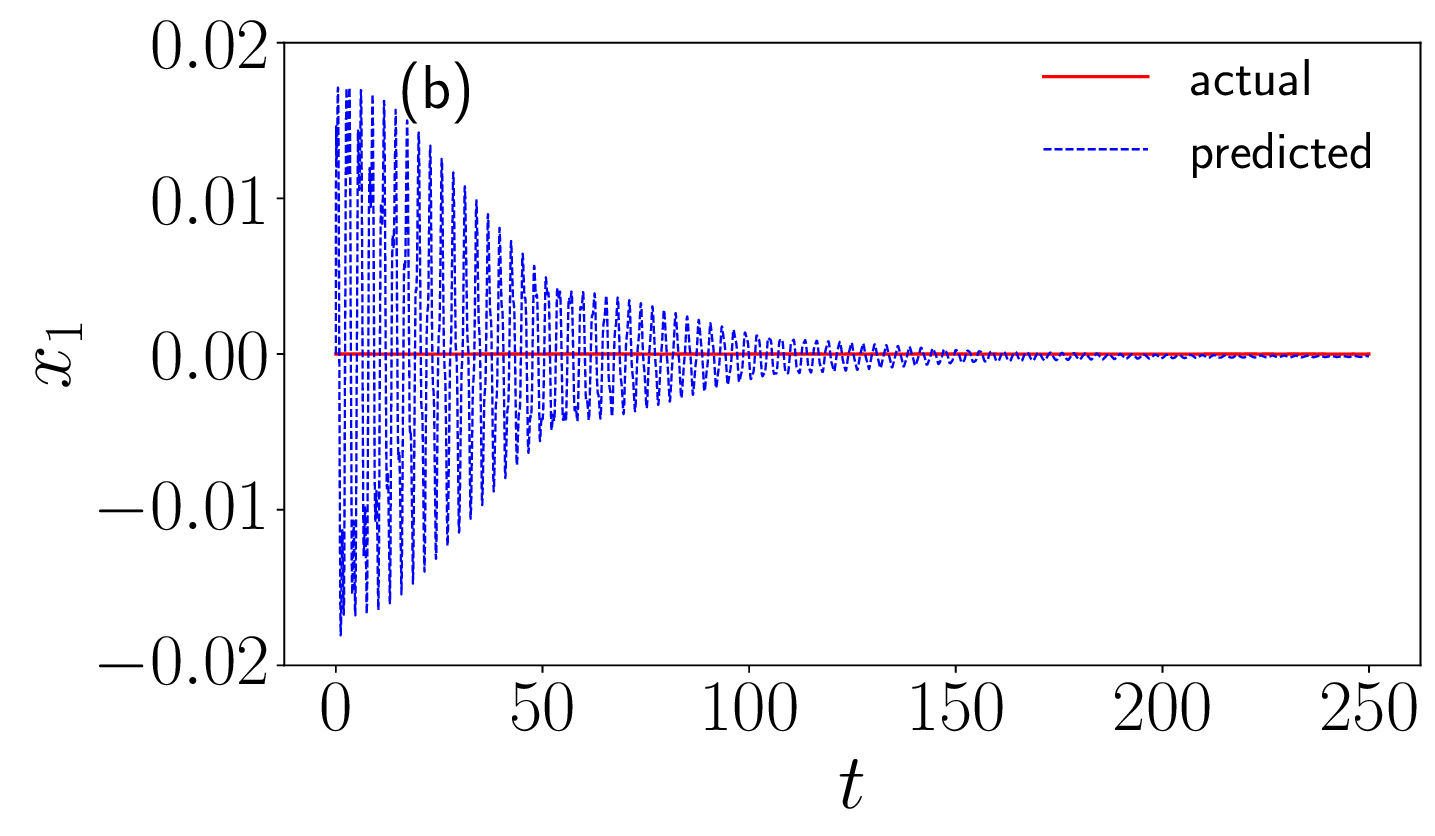}

\end{tabular}
\endgroup
\caption{Comparison of actual and predicted data of the coupled Stuart-Landau system in the torus regime,(a) $\varepsilon = 0.98$, (b) $\varepsilon = 1.02$. Other parameters are ($m = 300, b = 2, \alpha = 1, \rho = 0.18, \sigma = 0.01, k_{\varepsilon} = 0.52, \varepsilon_b = 0$). The RC was trained at $\varepsilon = 0.85, 0.9, 0.95$.} 
\label{fig:8}
\end{figure}

We further probe the mechanism of torus generation in the trained map. For the $\varepsilon = 1$ case, after the first conjugate pair crosses the unit circle, the ($x_1, y_1$) phase space still has an invariant curve; however, it only exists until the second pair crosses. Once both the pairs are outside the unit circle, the invariant curve is destroyed and a torus is generated in the phase plane. We perform Fourier analysis on all the four time series ($x_1(t), x_2(t), y_1(t), y_2(t)$) for both the cases (Fig.~\ref{fig:3}). For a limit cycle existing in the phase space ($\varepsilon = 3.6$), each time series contains a single common frequency. Conversely, when a torus exists in the phase space, each time series comprises two frequencies. In both cases ($\varepsilon = 3.6, \varepsilon = 1$), for different random realizations of $W_b, W_{in}, \mathcal{A}$, a different map is produced. Thus, to make the correct predictions, the time series generated by the map should also be composed of the same frequencies for different random realizations. This is achieved by keeping the arguments ($\theta = tan^{-1}(Im(\mu)/Re(\mu))$) of the eigenvalues that cross the unit circle unchanged. In other words, for all the different maps obtained after a successful training, the angle made by the eigenvalues from the real axis in the argand plane remains the same.

When $\varepsilon = 3.6$, assuming that $\mu = e^{\pm i\theta(0)}$ is the eigenvalue pair crossing the unit circle, the value of $\theta(0)/ 2\pi = 0.0358$. This value closely matches the frequency identified by the Fourier analysis of the original systems time series. At the bifurcation point ($\varepsilon = 1$), assuming the eigenvalues that cross are $\mu_1 = e^{\pm i\theta_1}$ and $\mu_2 = e^{\pm i\theta_2}$, $\theta_1/ 2\pi = 0.0174$ and $\theta_2/ 2\pi = 0.0541$. These values too, closely match the frequencies isolated by the Fourier analysis of the original system's time series. These observations can be explained using the normal form of the Neimark-Sacker bifurcation (discussed in detail in the supplementary material \cite{supp}). 

We also check the implications of a poorly trained machine. Training the same system for a low dimensional reservoir (having 50 nodes) and keeping all the other parameters unchanged, it is observed that the machine is no longer able to predict correctly. The corresponding analysis of map also reveals that a Neimark-Sacker bifurcation  no longer takes place (see supplementary material \cite{supp} for figures corresponding to a poorly trained reservoir).

We consider another example of Van der Pol oscillators coupled through mean field diffusive coupling \cite{PhysRevE.89.062902}:
\begin{eqnarray}\label{eq6}
     &&\dot x_{i} = y_{i} + \varepsilon(Q\Bar{X} - x_i)\\
    &&\dot y_{i} = a(1 - x_{i}^2)y_{i} - x_i.
\end{eqnarray}
We fix the parameters $a = 0.35$ and $Q = 0.5$, while the coupling strength $\varepsilon$ is varied. Here, $\Bar{X}$ is the mean field of the system. At $\varepsilon = 0.7$, there exists a Hopf bifurcation in the system. For $\varepsilon<0.7$, the system is in oscillatory state, while for $\varepsilon>0.7$, there exists a stable fixed point leading to amplitude death.
To learn this dynamics, the dimension of the reservoir was chosen as $600$. After successful training in the oscillatory region and generation of the corresponding reservoir map, the Jacobian analysis was repeated near the bifurcation parameter. For values of $\varepsilon$ greater than $0.7$, all eigenvalues are inside the unit circle, and most of them appear in complex conjugate pairs. At the bifurcation point, similar to the Stuart-Landau case discussed above, a single pair of complex conjugate eigenvalues crosses the unit circle, which indicates a Neimark-Sacker bifurcation. This is how the learned reservoir map imitates the limit cycle.

Fourier analysis of all the four time series ($x_1(t),x_2(t),y_1(t),y_2(t)$), reveal that all are composed of a single frequency, which is $0.15\hspace{0.1cm} Hz$. Assuming that the eigenvalues which cross are $\mu = e^{\pm i \theta}$ at the bifurcation point $(\epsilon = 0.7)$, the value of $\theta/2\pi$ is nearly equal to $0.00141$, again comes very close to the dominant frequency in the Fourier transform (see \cite{supp}).

\paragraph{Conclusion:}
Fundamentally, RC functions as a discrete map system. Our study illustrates how applying dynamical concepts improves the understanding of the prediction capabilities of the RC framework. Our numerical analysis demonstrates how the trained reservoir map successfully generates the dynamics of a continuous system. To learn the correct dynamical behavior around the Hopf bifurcation, the map undergoes a Neimark-Sacker bifurcation. It also correctly captures the intricate behaviour in the phase space like torus formation by the successive crossing of two conjugate eigenvalue pairs. The method used in this letter for the analysis of parameter-aware RC algorithm can be extended to other RC architectures like the single node reservoir \cite{Appeltant2011}, next generation reservoir computing \cite{Gauthier2021} to understand their learning mechanisms. Continuing in a similar vein, the mechanism for techniques such as crisis prediction learning \cite{PhysRevResearch.3.013090} can be explored as a question for the future.

\begin{acknowledgments}
SJ and DS gratefully acknowledge SERB Power grant SPF/2021/000136 and CSIR fellowship under award no. CSIRAWARD/JRF-NET2024/14347, respectively. The authors also thank the complex
systems lab members at IIT Indore for fruitful discussions.
\end{acknowledgments}

\nocite{*}


\end{document}


\preprint{APS/123-QED}
\title{Supplemental material for ``Dynamical analysis of a parameter aware Reservoir Computer''}

\author{Dishant Sisodia}
\email{dishantsisodia.physics@gmail.com}
\affiliation{Complex Systems Lab, Department of Physics, Indian Institute of
Technology Indore, Khandwa Road, Simrol, Indore-453552, India}
\author{Sarika Jalan}%
 \email{sarika@iiti.ac.in}
\affiliation{Complex Systems Lab, Department of Physics, Indian Institute of
Technology Indore, Khandwa Road, Simrol, Indore-453552, India}
\maketitle

\section{Neimark-Sacker Bifurcation in discrete time systems}{\label{sec1}}

Suppose that there is a generic discrete time, two-dimensional system of the form $ x_{n+1} = f(x_n, \alpha)$ and for sufficiently small $|\alpha|$, the fixed point is $x_0 = 0$. Let the complex multipliers be $\mu_{1,2}(\alpha) = r(\alpha)e^{\pm i\theta(\alpha)}$. If $r(0) = 1$ at $x_0 = 0$, then there exists a unique closed invariant curve near the fixed point assuming the following conditions hold (further analysis and proofs can be found in \cite{Kuznetsov2023}\cite{Kuznetsov:2008}):
\begin{eqnarray}\label{eqA1}
    &&r'(0) \neq 0 \\
    &&e^{ik\theta(0)} \neq 1 \quad \quad \text{for}\quad  k = 1,2,3,4.
\end{eqnarray}
The system is said to undergo a Neimark-Sacker bifurcation at $\alpha = 0, x_0 = 0$. Since it is very similar to the Hopf bifurcation in the continuous systems, it is often known as `Hopf bifurcation for maps' in the literature.
Any such discrete time, two dimensional system can be transformed into the following normal form:
\begin{equation}\label{eqA2}
\begin{aligned}
    &\begin{pmatrix}
        y_1 \\
        y_2
    \end{pmatrix}\rightarrow  (1 + \beta) \begin{pmatrix}
        cos \theta(\beta) & -sin \theta(\beta) \\
        sin \theta(\beta) & cos \theta(\beta)
    \end{pmatrix} \begin{pmatrix}
        y_1 \\
        y_2
    \end{pmatrix} + (y_{1}^2 + y_{2}^2)\begin{pmatrix}
        cos \theta(\beta) & -sin \theta(\beta) \\
        sin \theta(\beta) & cos \theta(\beta)
    \end{pmatrix} \begin{pmatrix}
        a(\beta) & -b(\beta) \\
        b(\beta) & a(\beta)
    \end{pmatrix}\begin{pmatrix}
        y_1 \\
        y_2
    \end{pmatrix}+ \mathcal{O}(\| y^4 \|),
\end{aligned}
\end{equation}
where $r(\alpha)$ is written as $1 + \beta(\alpha)$. If the higher order terms are negligible, then it can be transformed to polar co-ordinates to yield \cite{Kuznetsov2023}:
\begin{equation}\label{eqA3}
    \begin{aligned}
        &\rho \rightarrow \rho(1 + \alpha + a(\alpha)\rho^2) + \rho^{4}R_{\alpha}(\rho)\\
        &\phi \rightarrow \phi + \theta(\alpha) + \rho^{2}Q_{\alpha}(\rho),
    \end{aligned}
\end{equation} where $R, Q$ are some smooth functions. If we have $a(0) < 0$, then for the $\rho$ map, apart from the trivial fixed point at $\rho = 0$, there is another fixed point for $\alpha>0$:

\begin{equation}
    \rho_{0}(\alpha) = \sqrt{-\dfrac{\alpha}{a(\alpha)}} + \mathcal{O}(\alpha),
\end{equation}
which corresponds to the invariant curve due to the Neimark-Sacker bifurcation. The $a(0) > 0$ is analyzed in a similar fashion. The $\phi$ map in Eq. \ref{eqA3} describes a rotation which is approximately equal to $\theta(0)$ near the bifurcation point. Therefore, if the ratio $\theta(0)/2\pi$ is rational, then all the orbits on invariant circle are periodic with the frequency equal to $\theta(0)/2\pi$. Otherwise, if the ratio is irrational, all the orbits densely fill the invariant circle.

\section{Dimensional reduction using center manifold theorem}

To generalize the concept of Neimark-Sacker bifurcation to a $m$-dimensional system, we can use the center manifold theorem (details of this technique can be found in \cite{Wiggins2003}) to effectively reduce the system to two dimensions. Each system has three invariant subspaces $E^s, E^u, E^c$, corresponding to the span of generalized eigenvectors of the system's Jacobian matrix. The stable and unstable subspaces, $E^s$ and $E^u$, are spanned by the eigenvectors associated with eigenvalues whose modulus are less than and greater than 1, respectively. The center subspace, $E^c$, is spanned by the remaining eigenvectors, which correspond to eigenvalues with modulus equal to 1. If the $E^u$ subspace is null (no eigenvalue has modulus greater than 1), then all trajectories will rapidly decay to $E^c$. Thus, the long term behaviour of trajectories is determined by the center manifold $E^c$. If there exists a $m$ dimensional system spanned by $r_1,....,r_m$ having $m-2$ eigenvalues with modulus less than 1 and two eigenvalues having modulus equal to 1, then the system has a stable manifold $E^s$ ($\mathbb{R}^{m-2}$) and the system's dynamics can be restricted to the center manifold $E^c$ ($\mathbb{R}^2$). Hence, the generalization to a multi-dimensional case follows if the conditions in Eq. \ref{eqA1} are satisfied. In that case, an invariant curve will arise in the center manifold ($E^c$) spanned by $\tilde {(r_{1})}, \tilde{(r_{2})}$ and the remaining eigenvectors of $E^s$ will decay to 0 after sufficient transient time. This implies that $\tilde{(r_{1})}$ and $\tilde{(r_{2})}$ have a common period. Since each element of the original basis ($r_1,....,r_m$), is just a linear combination of the eigen basis $\tilde{(r_{1})},....,\tilde{(r_{m})}$, it implies that after removing sufficient transient, each $r_{i}$ will have the same common period. Therefore, any $r_{i}, r_{j}$ pair will also form an invariant curve in the phase space, which explains the observation presented in the main text.

We conclude the existence of a Neimark-Sacker bifurcation in our system by checking the conditions in Eq.~\ref{eqA1} numerically. Plotting the modulus of eigenvalues as a function of bifurcation parameter ensures that the first condition in Eq. \ref{eqA1} is satisfied. For the second condition, $\theta(0)$ is simply calculated as $tan^{-1}(Im(\mu)/Re(\mu))$ to be checked for $k = 1,2,3,4$.

\section{Supplementary figures for coupled Van der Pol Oscillators}
\subsection{Predictions:}
\begin{figure}[h!]
\begingroup
\begin{tabular}{c}
\includegraphics[scale=0.3]{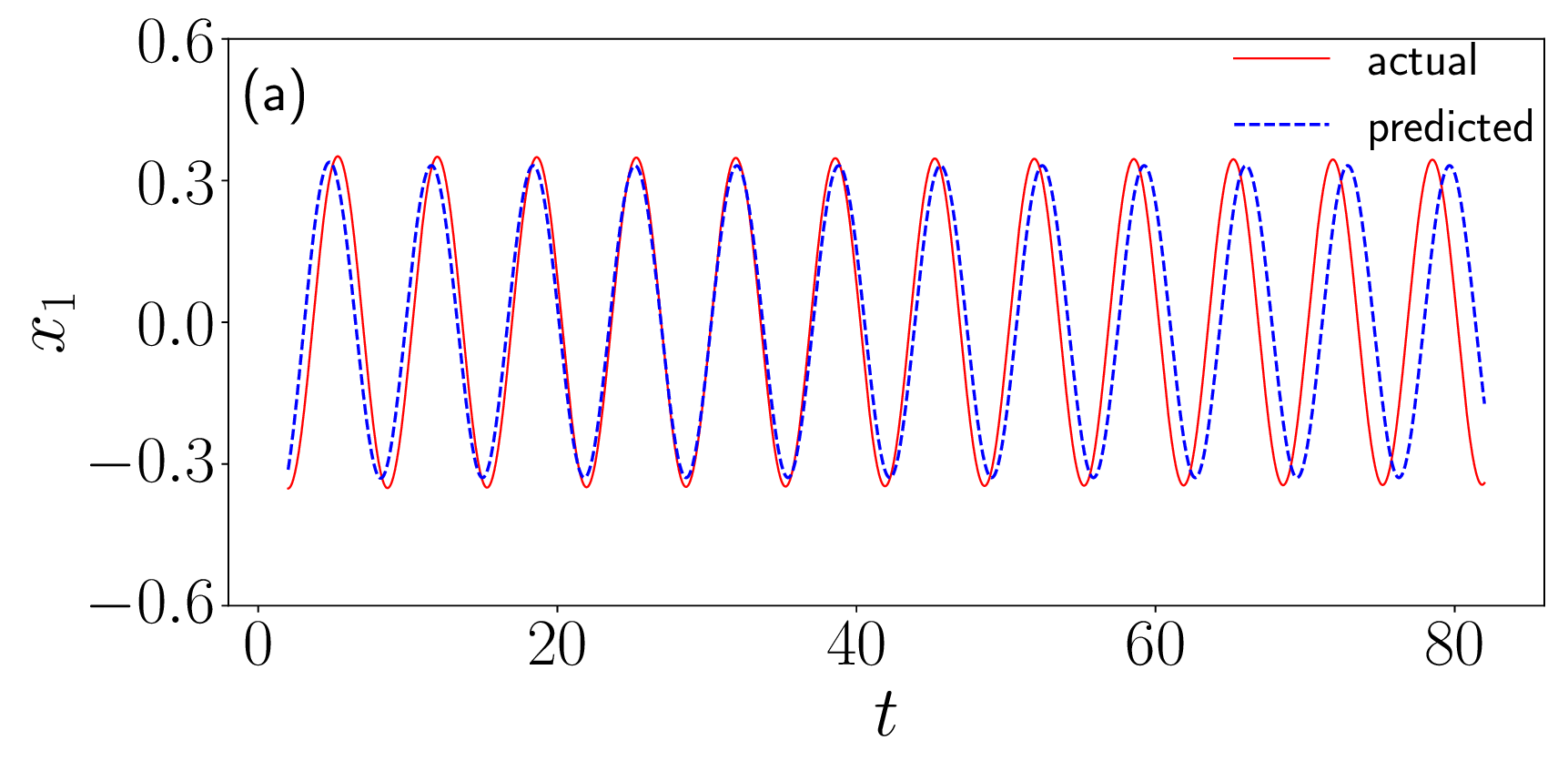}\\
\includegraphics[scale=0.3]{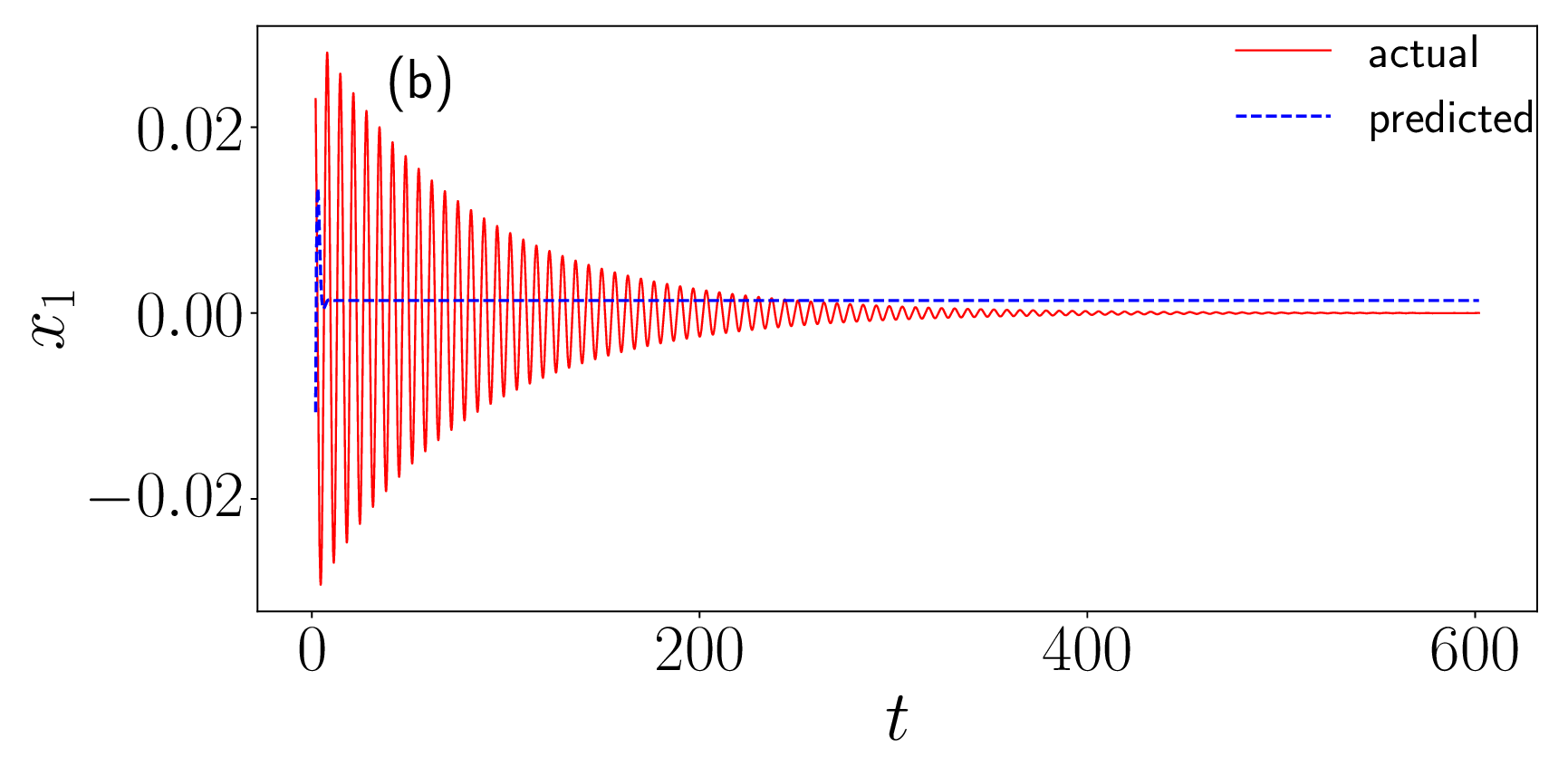}

\end{tabular}
\endgroup
\caption{(a) Comparison of actual and predicted data for $\varepsilon = 0.68$. (b) Comparison of actual and predicted data for $\varepsilon = 0.75$.}
\label{fig:S1}
\end{figure}

Fig.~\ref{fig:S1} shows the comparison between actual and predicted data in the system of coupled Van der Pol oscillators (Fig. 1) after a successful training of reservoir at $\varepsilon = 0.56, 0.61$ and $0.66$. The hyperparameters chosen for the training are m = 600, b = 2, $\alpha = 0.66$, $\rho = 0.18$, $\sigma = 0.03$, $k_{b} = 0.52$, $\varepsilon_b = 0$.

\subsection{Change in the eigenvalues with $\varepsilon$}

\begin{figure}[h!]
    \centering
    \includegraphics[width=0.5\textwidth]{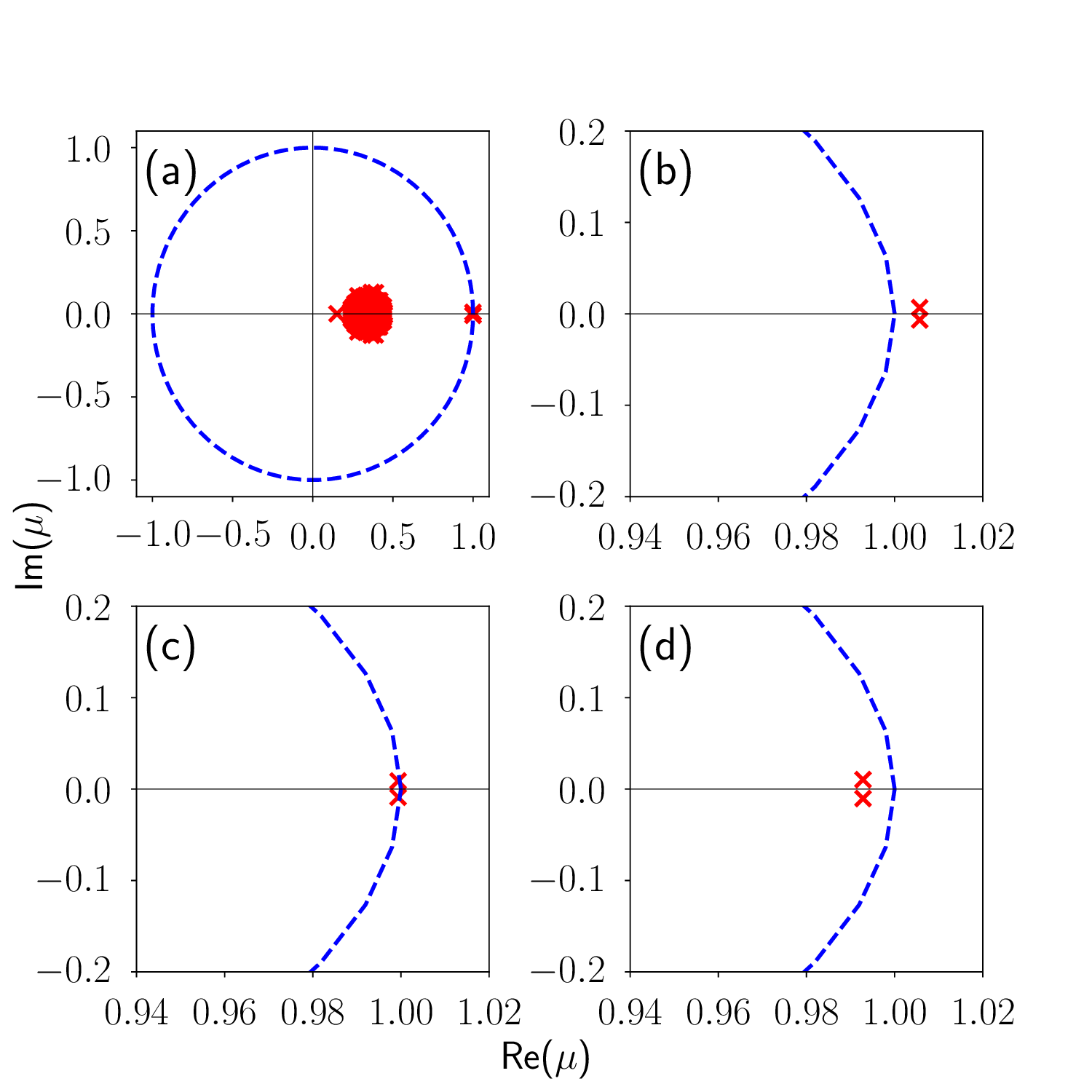}
    \caption{Behaviour of eigenvalues for coupled Van der Pol oscillators. (a) A single conjugate pair crosses the unit circle at $\varepsilon = 0.7$. (b), (c) and (d) show the eigenvalues at $\varepsilon = 0.66, 0.7$ and $0.74$ respectively.}
    \label{fig:S2}
\end{figure}

In Figure \ref{fig:S2}, the behaviour of the eigenvalues of the Jacobian matrix of the coupled Van der Pol oscillators is shown as the bifurcation parameter is varied. Only a single conjugate pair crosses the unit circle, indicating a Neimark-Sacker bifurcation. The value of $\theta/2\pi$ assuming that the eigenvalues which cross are $\mu = e^{\pm i \theta}$ at the bifurcation point $(\epsilon = 0.7)$ is nearly equal to $0.00141$.

\subsection{Fourier analysis of the time series}
\begin{figure}[h!]
    \centering
    \includegraphics[scale=0.3]{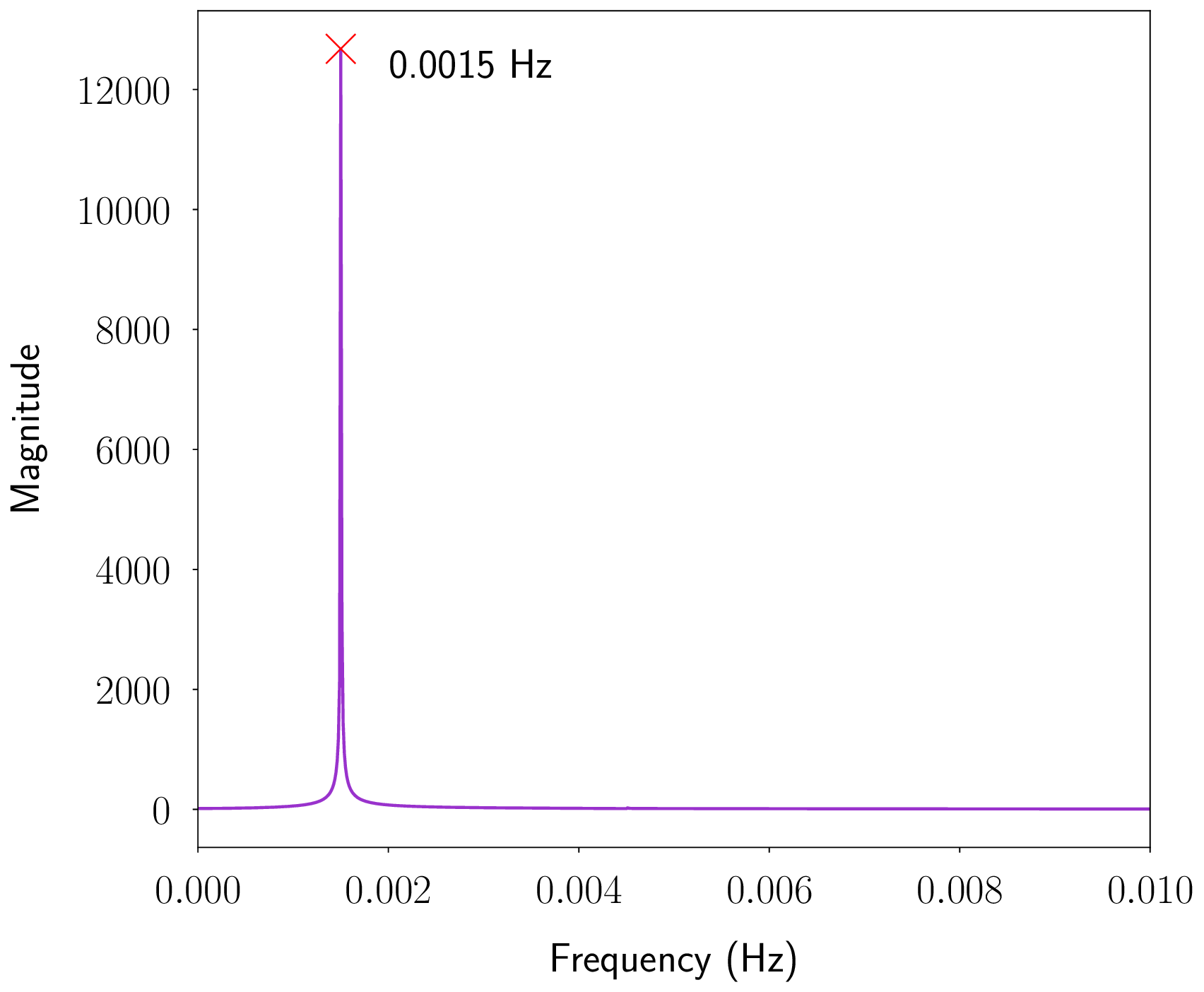}
    \caption{The dominant frequencies of the time series are isolated along with their magnitudes by performing the fourier transform of $x_1(t)$ for Van der Pol Oscillator at $\epsilon = 0.69$}
    \label{fig:S3}
\end{figure}

Performing the Fourier analysis on all four time series ($x_1(t),x_2(t),y_1(t),y_2(t) $) of the van der Pol oscillator system leads to the observation that all of them are composed of a single frequency (Fig. \ref{fig:S3}) which is $0.0015\hspace{0.1cm} Hz$. It is very close to the value of frequency obtained by the eigenvalue analysis.

\section{Analysis for poorly trained machine}
\begin{figure}[h!]
\begingroup
\begin{tabular}{c}
\includegraphics[scale = 0.35]{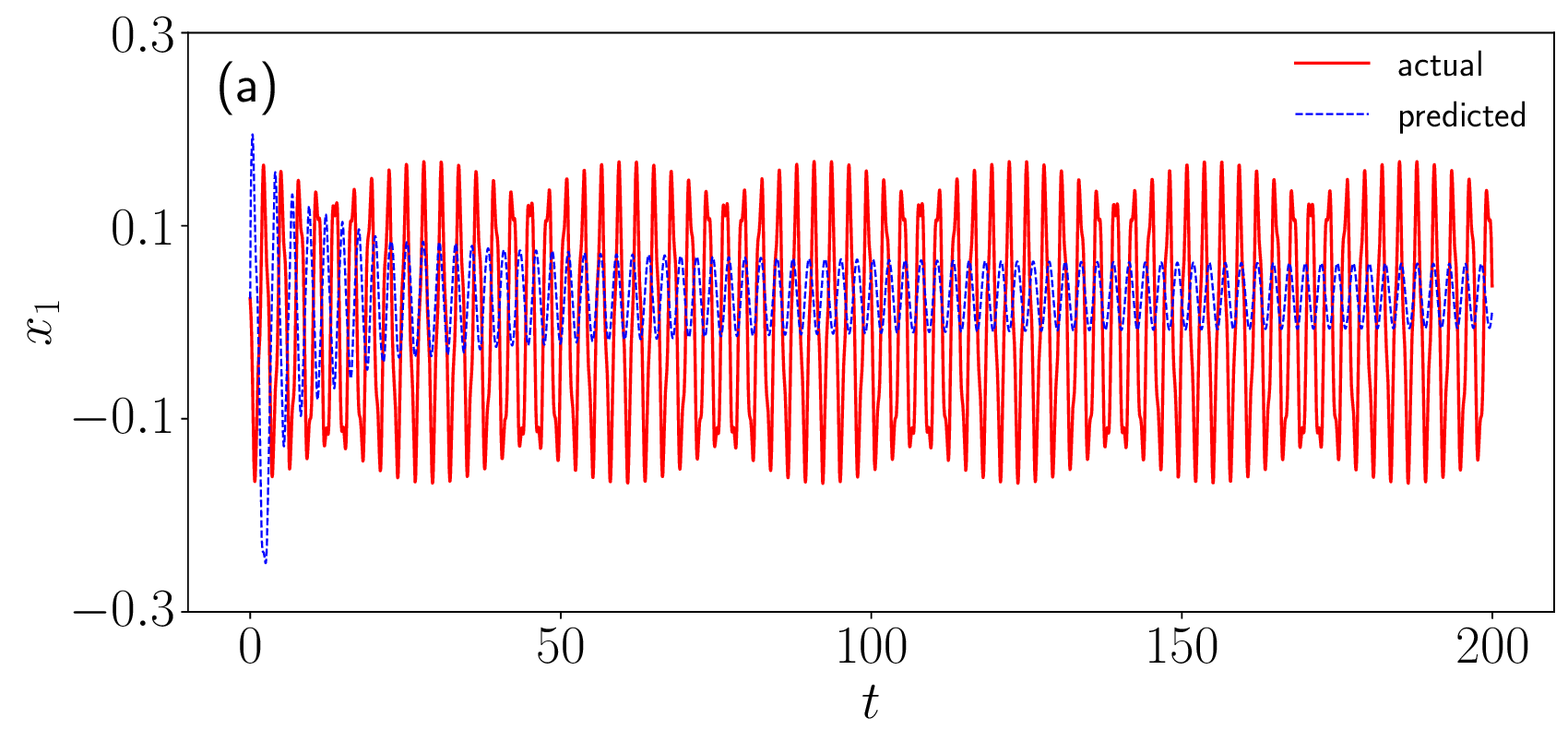}\\
\includegraphics[scale = 0.35]{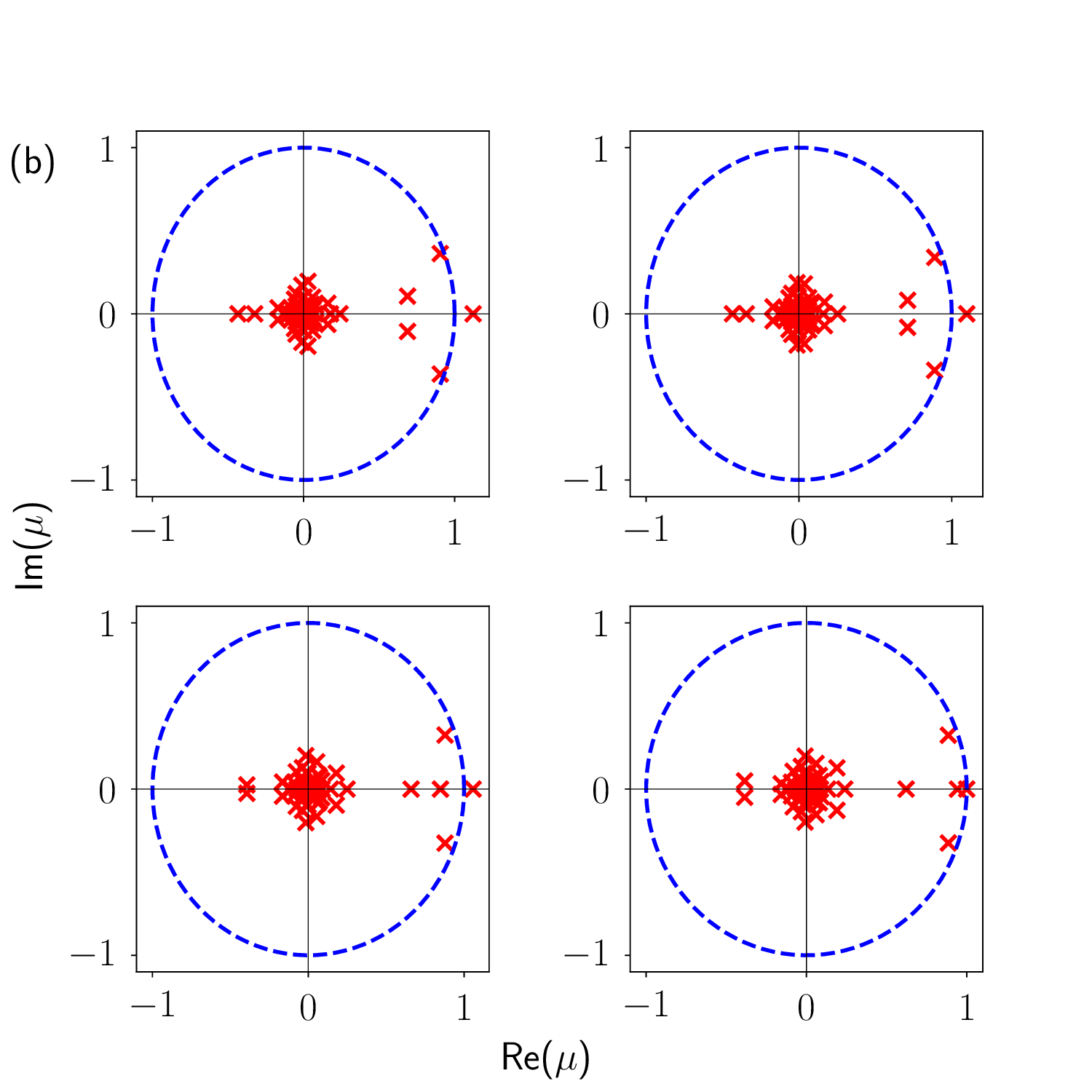}

\end{tabular}
\endgroup
\caption{(a) Comparison of actual and predicted data for $\varepsilon = 0.98$. (b) Motion of eigenvalues (at $\varepsilon = 0.94, 0.99, 1.04$ and $1.09$)}
\label{fig:S4}
\end{figure}

We reduced the dimension of the reservoir to 50 and observed that machine could not correctly predict the dynamics of coupled Stuart-Landau oscillators around $\varepsilon = 1$. All other hyperparameters apart from the dimension of the reservoir were kept constant. In order to predict the torus dynamics there is no longer a crossing of two complex conjugate pairs as shown in Fig. \ref{fig:S4}. Near the bifurcation parameter we can no longer conclude the existence of Neimark-Sacker bifurcation.

\nocite{*}